\documentclass[10pt]{article}
\usepackage{epsfig}
\renewcommand{\author}[2]{\begin{center}
                           {\sc #1}\\
                           {#2}
                          \end{center}}
\renewcommand{\title}[1]{\begin{center}
                      {\Large {\bf #1}}
                      \end{center}}
\def\be{\begin{equation}}
\def\ee{\end{equation}}
\def\lb{\left[}
\def\rb{\right]}

\begin{document}

\title{An exact Riemann Solver for multidimensional
special relativistic hydrodynamics
\vspace*{5mm}}

\author{Jos\'e A. Pons$^1$,  Jos\'e M$^{\underline{\mbox a}}$ Mart\'{\i}$^1$ 
and Ewald M\"uller$^2$
\vspace{5mm}}
              
{$^1$Departament d'Astronomia i Astrof\'{\i}sica,
U. Val\`encia, 46100 Burjassot, Spain \\
$^2$ MPI f\"ur Astrophysik, Karl-Schwarzschild-Str. 1,
85748 Garching, Germany}
\vspace{5mm}

e-mail:~jose.a.pons@uv.es

\vspace{5mm}

\begin{abstract}

  We have generalised the {\it exact} solution of the Riemann problem
in special relativistic hydrodynamics \cite{MM94}
for arbitrary tangential flow velocities. The solution is obtained by
solving the jump conditions across shocks plus an ordinary
differential equation arising from the self-similarity condition along
rarefaction waves, in a similar way as in purely normal flow. The
dependence of the solution on the tangential velocities is analysed.
This solution has been used
to build up an {\it exact Riemann solver} 
implemented in a multidimensional relativistic 
(Godunov-type) hydro-code. 

\end{abstract}

\section{Introduction}
\label{s:intro}

  The decay of a discontinuity separating two constant initial states
({\it Riemann problem}) has played a very important role in the
development of numerical hydrodynamic codes in classical (Newtonian)
hydrodynamics after the pioneering work of Godunov \cite{God59}. Nowadays,
most modern high-resolution shock-capturing methods \cite{Le92} are
based on the exact or approximate solution of Riemann problems between
adjacent numerical cells and the development of efficient Riemann
solvers has become a research field in numerical analysis in its own
(see, e.g., \cite{To98}).

  Riemann solvers began to be introduced in numerical relativistic
hydrodynamics at the beginning of the nineties and,
presently, the use of high-resolution
shock-capturing methods based on Riemann solvers is considered as the
best strategy to solve the equations of relativistic hydrodynamics in
nuclear physics (heavy ion collisions) and astrophysics (stellar core
collapse, supernova explosions, extragalactic jets, gamma-ray
bursts). This fact has caused a rapid development of Riemann solvers
for both special and general relativistic hydrodynamics \cite{IM99,MM99}.

The main idea
behind the solution of a Riemann problem (defined by two constant
initial states, $L$ and $R$, at left and right of their common contact
surface) is that the self-similarity of the flow through rarefaction
waves and the Rankine-Hugoniot relations across shocks allow one to
connect the intermediate states $I_*$ ($I = L,R$) with their
corresponding initial states, $I$.  The analytical solution of the
Riemann problem in classical hydrodynamics \cite{CF48}
rests on the fact that the normal velocity in the
intermediate states, $v^n_{I_*}$, can be written as a function of the
pressure $p_{I_*}$ in that state (and the flow conditions in state
$I$). Thus, once $p_{I_*}$ is known, $v^n_{I_*}$ and all other unknown
state quantities of $I_*$ can be calculated. 

  In the case of relativistic hydrodynamics the same procedure can be
followed \cite{MM94,PMM99}, the major difference with classical 
hydrodynamics stemming
from the role of tangential velocities. While in the classical case
the decay of the initial discontinuity does not depend on the
tangential velocity (which is constant across shock waves and
rarefactions), in relativistic calculations the components of the flow
velocity are coupled through the presence of the Lorentz factor. 

\section{The equations of relativistic hydrodynamics}
\label{s:eqs}

  The equations of relativistic hydrodynamics admit a conservative formulation
which has been exploited in the last decade to implement high-resolution 
shock-capturing methods. In Minkowski space time the equations in this 
formulation read 
\be 
\partial_t {\bf U} + \partial_i {\bf F}^{(i)} = 0
\label{cl}
\ee
where ${\bf U}$ and ${\bf F}^{(i)}({\bf U})$ ($i=1,2,3$) are, respectively, the
vectors of conserved variables and fluxes 
\be
{\bf U} = (D,S^1,S^2,S^3,\tau)^T
\ee
\be
{\bf F}^{(i)} = (D v^i, S^1 v^i + p \delta^{1i}, S^2 v^i + p \delta^{2i}, 
S^3 v^i + p \delta^{3i},S^i-D v^i)^T.
\ee
The conserved variables (the rest-mass density, $D$, the momentum density, 
$S^i$, and the energy density $\tau$) are defined in terms of the {\it 
primitive variables},$(\rho, v^i, \varepsilon)$, according to
\begin{eqnarray}
D    = \rho W   ~, \quad     
S^i  = \rho h W^2 v^i   ~, \quad  
\tau = \rho h W^2 - p - D
\end{eqnarray}
where $W=(1-v^2)^{-1/2}$ is the Lorentz factor
and $h = 1 + \varepsilon + p / \rho$ the specific enthalpy.

  In the following we shall restrict our discussion to an ideal gas 
equation of state with
constant adiabatic exponent, $\gamma$, for which the specific internal energy 
is given by
\be
\varepsilon = \frac{p}{(\gamma -1)\rho}.
\ee

\section{Relation between the normal flow velocity and pressure behind 
         relativistic rarefaction waves}
\label{s:raref}

Choosing the
surface of discontinuity to be normal to the $x$-axis, rarefaction
waves are self-similar solutions of the flow equations depending
only on the combination $\xi=x/t$. Getting rid of all the terms with $y$
and $z$ derivatives in equations (\ref{cl}) and substituting the
derivatives of $x$ and $t$ in terms of the derivatives of $\xi$, the
system of equations can be reduced to
just one ordinary differential equation (ODE) and two algebraic conditions
\begin{eqnarray}
\rho h W^2 (v^x-\xi) dv^x + (1-\xi v^x) dp = 0 \label{ode} \label{rar1} \\
h W v^y = \mbox{constant} \label{alg1} \label{rar2} \\
h W v^z = \mbox{constant} \label{alg2} \label{rar3},
\end{eqnarray}
with $\xi$ constrained by 
\begin{eqnarray}
\xi & = & \frac{v^x(1-c_s^2) \pm c_s
\sqrt{(1-v^2)[1 - v^2 c_s^2 - (v^x)^2(1-c_s^2)]}}{1-v^2c_s^2},
\label{xi6} 
\end{eqnarray}
because non-trivial similarity solutions exist only if the Wronskian
of the original system vanish. 
We have denoted by $c_s$ the speed of sound, provided by the equation of state.
The plus and minus sign correspond to rarefaction waves propagating
to the right (${\cal R}_{\rightarrow}$) and
left (${\cal R}_{\leftarrow}$), respectively.  The two
solutions for $\xi$ correspond to the maximum and minimum eigenvalues
of the Jacobian matrix associated to ${\bf F}^{(x)}({\bf U})$ \cite{IM99,MM99}, 
generalizing the result found for
vanishing tangential velocity \cite{MM94}.

From equations (\ref{rar2}) and
(\ref{rar3}) it follows that $v^y/v^z =$ constant, i.e., the
tangential velocity does not change direction across rarefaction waves.
Notice that, in a
kinematical sense, the Newtonian limit ($v^{i} \ll 1$) leads to $W=1$,
but equations (\ref{rar2}) and (\ref{rar3}) do not reduce to the
classical limit $v^{y,z}=$ constant, because the specific enthalpy
still couples the tangential velocities. Thus, even for slow flows,
the Riemann solution presented in this paper must be employed for
thermodynamically relativistic situations ($h \gg 1$).  The same
result can be deduced from the Rankine-Hugoniot relations for shock
waves (see next section).

  Using (\ref{xi6}) and the definition of the sound speed,
the ODE (\ref{ode}) can be written as
\be
\frac{dv^x}{dp}= \pm \frac{c_s}{W^2 \gamma p}
\frac{1}{\sqrt{1+g(\xi_{\pm},v^x,v^t)}}
\label{ode2}
\ee
where $v^t=\sqrt{(v^y)^2 + (v^z)^2}$ is the absolute value of the tangential 
velocity and 
\be
g(\xi_{\pm},v^x,v^t)=\frac{(v^t)^2 (\xi_{\pm}^2-1)}{(1-\xi_{\pm} v^x)^2}.
\ee

  Considering that in a Riemann problem the state ahead of the
rarefaction wave is known, equation (\ref{ode2}) can be integrated
with the constraint  $h W v^t = $ constant, allowing 
to connect the states ahead ($a$) and behind ($b$) the rarefaction
wave. The solution is only a function of
$p_b$ and can be stated in compact form as
\be
v^x_b = {\cal R}^a_{\rightleftharpoons}(p_b).
\ee

\section{Relation between post-shock flow velocities and pressure
for relativistic shock waves.}

  The Rankine-Hugoniot conditions relate the states on both sides of a shock 
and are based on the continuity of the mass flux and the energy-momentum flux 
across shocks.  Their relativistic version was first obtained by Taub 
\cite{Tau48} (see also \cite{Tau78,Koe80}).

  Considering the surface of discontinuity as normal to the 
$x$-axis,  the invariant mass flux across the shock can be written as
\be
j \equiv W_s D_a (V_s-v^x_a) = W_s D_b (V_s-v^x_b).
\label{mflux}
\ee
where $V_s$ is the coordinate velocity of the hyper-surface
that defines the position of the shock wave and $W_s$ is the 
correspondent Lorentz factor, $W_s = (1-V_s^2)^{1/2}$.
According to our definition, $j$ is positive for shocks propagating to
the right. 
In terms of the mass flux, $j$, the Rankine-Hugoniot conditions are
\be
[v^x] = -\frac{j}{W_s} \lb \frac{1}{D} \rb,
\label{vx}  
\ee
\be
[p] = \frac{j}{W_s} \lb \frac{S^x}{D} \rb,  
\label{p}
\ee
\be
\lb {h W v^y} \rb = 0, 
\label{rhvy} 
\ee
\be
\lb {h W v^z} \rb = 0, 
\label{rhvz} 
\ee
\be
[v^x p] = \frac{j}{W_s} \lb \frac{\tau}{D} \rb. 
\label{vp}
\ee
Equations~(\ref{rhvy}) and (\ref{rhvz}) imply that the quantity
$hWv^{y,z}$ is constant across a shock wave and, hence, that the
orientation of the tangential velocity does not change. The latter
result also holds for rarefaction waves (see \S3). Equations
(\ref{vx}), (\ref{p}) and (\ref{vp}) can be
manipulated to obtain $v^x_b$ as a function of $p_b$,
$j$ and $V_s$. Using the relation $S^x=(\tau+p+D)v^x$ and after some
algebra, one finds
\be
v^x_b = \left( h_a W_a v^x_a + \frac{W_s (p_b-p_a)}{j} \right)
\left( h_a W_a + (p_b - p_a) \left(\frac{W_s v^x_a}{j} +
\frac{1}{\rho_a W_a} \right) \right)^{-1}.
\ee

  The final step is to express $j$ and $V_s$ as a function of the post-shock 
pressure. First, from the definition of the mass flux we obtain
\begin{equation}
\displaystyle{
V_s^{\pm} = \frac{ \rho_a^2 W_a^2 v^x_a  \pm   
                   |j| \sqrt{j^2 + \rho_a^2 W_a^2 (1 -{v_a^x}^2)}
            }{ \rho_a^2 W_a^2 + j^2 }
}
\label{velshock}
\end{equation}
where $V_s^{+}$ ($V_s^{-}$) corresponds to shocks propagating to the
right (left). 

Second, from the Rankine-Hugoniot relations
and the physical solution of $h_b$ obtained from the Taub adiabat
\cite{Tho73} (the relativistic version of the Hugoniot 
adiabat), that relates only
thermodynamic quantities on both sides of the shock, 
the square of the mass flux $j^2$ can be obtained as a function of $p_b$. 
Using the positive (negative) root
of $j^2$ for shock waves propagating towards the right (left),
the desired relation between
the post-shock normal velocity $v^x_b$ and the post-shock pressure
$p_b$ is obtained. In a compact way the relation reads
\be
v^x_b = {\cal S}^a_{\rightleftharpoons}(p_b).
\ee
We refer to the interested reader to references \cite{MM94,PMM99}
for further details.

\section{The solution of the Riemann problem with arbitrary
tangential velocities.}

The time evolution of a Riemann problem can be represented as:
\be
I\; \rightarrow \;L\;{\cal W}_{\leftarrow}\;L_*\;{\cal C}\;
R_*\;{\cal W}_{\rightarrow}\;R
\ee
where $\cal W$ denotes a simple wave (shock or rarefaction),
moving towards the initial left ($\leftarrow$) or right 
($\rightarrow$) states.
Between them, two new states appear, namely $L_*$ and $R_*$,
separated from each other through the third wave $\cal C$, which is a contact
discontinuity. Across the contact
discontinuity pressure and normal velocity are constant, while the
density and the tangential velocity exhibits a jump. 

  As in the Newtonian case, the compressive character of shock waves (density 
and pressure rise across the shock) allows us to discriminate between shocks 
($\cal S$) and rarefaction waves ($\cal R$):
\be
{\cal W}_{\leftarrow\;(\rightarrow)} =
\left\{ \begin{array}{rcl} {\cal R}_{\leftarrow\;(\rightarrow)}
&,& p_b \leq p_a\\
{\cal S}_{\leftarrow\;(\rightarrow)}
&,& p_b > p_a \end{array}
\right.
\label{shandrar}
\ee
where $p$ is the pressure and subscripts $a$ and $b$ denote quantities
ahead and behind the wave. For the Riemann problem $a\equiv L(R)$ and
$b \equiv L_* (R_*)$ for ${\cal W}_{\leftarrow}$ and ${\cal
W}_{\rightarrow}$, respectively. 

  The solution of the Riemann problem consists in finding the
positions of the waves separating the four states and the
intermediate  states, $L_*$ and $R_*$.
The functions ${\cal W}_{\rightarrow}$ and ${\cal
W}_{\leftarrow}$ allow one to determine the functions $v^x_{R*}(p)$
and $v^x_{L*}(p)$, respectively. The pressure $p_*$ and the flow
velocity $v^x_*$ in the intermediate states are then given by the
condition
\be
v^x_{R*}(p_*) = v^x_{L*}(p_*) = v^x_*.
\label{vp0}
\ee
which is an implicit algebraic equation in $p_*$ and
can be solved by means of an iterative method. 
When $p_*$ and $v^x_*$ have been obtained,
the equation of state gives the specific internal energy and the
remaining state variables of the intermediate state $I_*$ can be
calculated using the relations between $I_*$ and the respective
initial state $I$ given through the corresponding wave. 

\begin{figure}
\centerline{\epsfig{figure=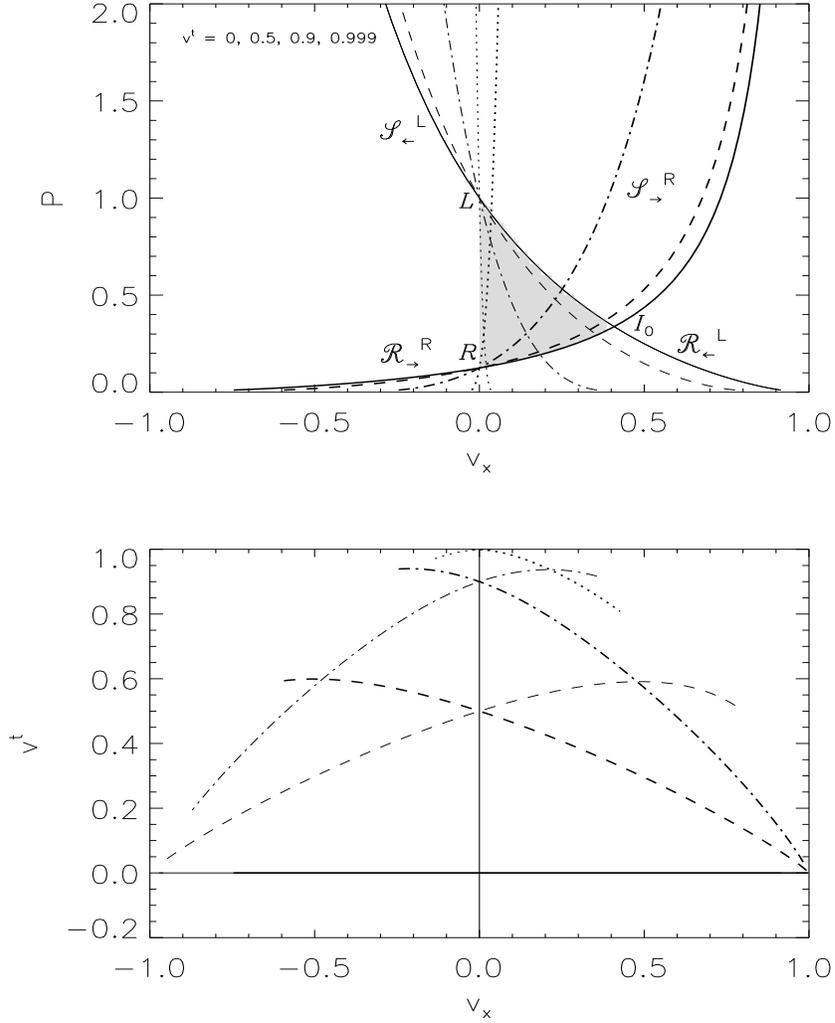,width=12cm,height=14.5cm}}
\caption{ Graphical solution in the $(p,v^x)$--plane (upper
panel) and in the $(v^t,v^x)$--plane (lower panel) of the relativistic
Riemann problem with initial data $p_L=1.0$, $\rho_L=1.0$,
$v^x_L=0.0$; $p_R=0.1$, $\rho_R=0.125$ and $v^x_R=0.0$ for different
values of the tangential velocity $v^t= 0, 0.5, 0.9, 0.999$,
represented by solid, dashed, dashed-dotted and dotted lines,
respectively. An ideal gas EOS with $\gamma=1.4$ was assumed.  The
crossing point of any two lines in the upper panel gives the pressure
and the normal velocity in the intermediate states. The value of the
tangential velocity in the states $L_*$ and $R_*$ is obtained from the
value of the corresponding functions at $v^x$ in the lower panel, and
$I_0$ gives the solution for vanishing tangential velocity.  The range
of possible solutions is given by the shaded region in the upper
panel.}
\label{fig3}
\end{figure}

Notice that the
solution of the Riemann problem depends on the modulus of $v^t$, but
not on the direction of the tangential velocity.  Figure \ref{fig3} shows 
the solution of a particular Riemann problem \cite{Sod78} for different
values of the tangential velocity $v^y=0, 0.5, 0.9, 0.99$.  The
crossing point of any two lines in the upper panel gives the pressure
and the normal velocity in the intermediate states.  
Whereas the pressure in the intermediate state can take any
value between $p_L$ and $p_R$, the normal flow velocity can be
arbitrarily close to zero in the case of an extremely relativistic
tangential flow. 

  To study the influence of tangential velocities on the solution a
Riemann problem, we have calculated the solution of a standard test
involving the propagation of a relativistic blast wave produced by a
large jump in the initial pressure distribution
for different combinations of tangential velocities \cite{PMM99}.
Although the structure of the solution remains unchanged for different
tangential velocities, the values
in the constant state may change by a large amount. 

\begin{figure}
\centerline{\epsfig{figure=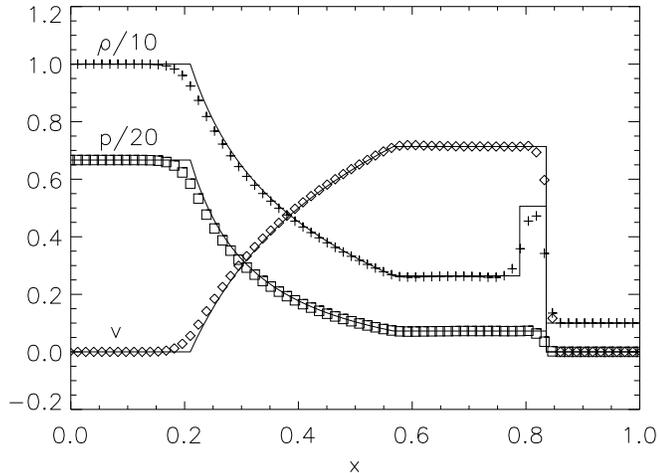,width=10cm}}
\caption{Exact (solid lines) and numerical profiles along
the diagonal of pressure (squares),
density (crosses) and normal velocity (diamonds) for the 
2--dimensional relativistic shock tube discussed in the text.  
}
\end{figure}

\section{Discussion and conclusions.}

  We have obtained the exact solution of the Riemann problem in
special relativistic hydrodynamics with arbitrary tangential
velocities. Unlike in Newtonian hydrodynamics, tangential velocities
are coupled with the rest of variables through the Lorentz factor,
present in all terms in all equations. It strongly affects the
solution, especially for ultra-relativistic tangential flows. In
addition, the specific enthalpy also acts as a coupling factor and
modifies the solution for the tangential velocities in
thermodynamically relativistic situations (energy density and pressure
comparable to or larger than the proper rest-mass density), rendering
the classical solution incorrect in slow flows with very large
internal energies.

  Our solution has interesting practical applications. First, it can
be used to test the different approximate relativistic Riemann
solvers and the multi-dimensional hydrodynamic codes based on directional
splitting.
Second, it can be used to construct multi-dimensional relativistic
Godunov-type hydro codes. 
As an example, we have simulated a relativistic 
tube \cite{Sod78}, in a $100 \times 100$ Cartesian grid, where the
initial discontinuity was located in a main diagonal. The initial states
were $\rho_L=10$,  $\rho_R=1$, $p_L=13.3$, $p_R=0.66E-3$, $v_L=0$, $v_R=0$,
and the adiabatic index is $\gamma=5/3$.
Spatial order
of accuracy was set to second order by means of a
monotonic piecewise linear reconstruction procedure and second order in
time is obtained by using a Runge-Kutta method for time advancing.
The exact solution of the Riemann problem is used at every interface
to calculate the numerical fluxes. The results are shown in Figure 2, 
and are comparable to those obtained with other HRSC methods. 
Profiles of all variables are stable and discontinuities are well resolved
without excessive smearing.
An efficient implementation of this exact Riemann solver in the context
of multidimensional relativistic PPM \cite{MM96} is in progress and
will be reported elsewhere.


\end{document}